\documentclass[aps,prb,10pt,twocolumn,superscriptaddress]{revtex4-1}

\usepackage{lmodern}
\usepackage[T1]{fontenc}
\usepackage[utf8]{inputenc}
\usepackage{graphicx}
\usepackage{amsmath}
\usepackage[unicode=true,
            colorlinks=true,
            linkcolor=blue,
            citecolor=blue,
            urlcolor=blue]{hyperref} 
\usepackage{siunitx}
\usepackage{multirow}

\renewcommand{\Re}{\operatorname{Re}}
\renewcommand{\Im}{\operatorname{Im}}

\renewcommand{\d}[1]{\ensuremath{\mathop{{\rm d}{#1}}}}

\begin{document}

\title{Thermoelectric detection of propagating plasmons in graphene}

\author{Mark B. Lundeberg}
\thanks{These authors contributed equally}
\affiliation{ICFO --- Institut de Ciències Fotòniques, The Barcelona Institute of Science and Technology, 08860 Castelldefels (Barcelona), Spain}
\author{Yuanda Gao}
\thanks{These authors contributed equally}
\affiliation{Department of Mechanical Engineering, Columbia University, New York, NY 10027, USA}
\author{Achim Woessner}
\affiliation{ICFO --- Institut de Ciències Fotòniques, The Barcelona Institute of Science and Technology, 08860 Castelldefels (Barcelona), Spain}
\author{Cheng Tan}
\affiliation{Department of Mechanical Engineering, Columbia University, New York, NY 10027, USA}
\author{Pablo Alonso-González}
\affiliation{CIC nanoGUNE, E-20018, Donostia-San Sebastián, Spain}
\author{Kenji Watanabe}
\affiliation{National Institute for Materials Science, 1-1 Namiki, Tsukuba 305-0044, Japan}
\author{Takashi Taniguchi}
\affiliation{National Institute for Materials Science, 1-1 Namiki, Tsukuba 305-0044, Japan}
\author{James Hone}
\affiliation{Department of Mechanical Engineering, Columbia University, New York, NY 10027, USA}
\author{Rainer Hillenbrand}
\affiliation{CIC nanoGUNE and EHU/UPV, E-20018, Donostia-San Sebastián, Spain}
\affiliation{IKERBASQUE, Basque Foundation for Science, 48011 Bilbao, Spain}
\author{Frank H. L. Koppens}
\email{frank.koppens@icfo.eu}
\affiliation{ICFO --- Institut de Ciències Fotòniques, The Barcelona Institute of Science and Technology, 08860 Castelldefels (Barcelona), Spain}
\affiliation{ICREA --- Institució Catalana de Recerça i Estudis Avancats, Barcelona, Spain.}

\date{\today{}}


\maketitle


{\bf
Controlling, detecting and generating propagating plasmons by all-electrical means is at the heart of on-chip nano-optical processing.\cite{Gramotnev2010,Vakil2011,Dyakonov1996}
Graphene carries long-lived plasmons that are extremely confined and controllable by electrostatic fields,\cite{Wunsch2006,Hwang2007,Jablan2009,Grigorenko2012} however electrical detection of propagating plasmons in graphene has not yet been realized. 
Here, we present an all-graphene mid-infrared plasmon detector, where a single graphene sheet serves simultaneously as the plasmonic medium and detector.
Rather than achieving detection via added optoelectronic materials, as is typically done in other plasmonic systems,
\cite{Ditlbacher2006,Neutens2009,Falk2009,Heeres2010,Dufaux2010,Goykhman2011,Goodfellow2015,Brongersma2015}
our device converts the natural decay product of the plasmon---electronic heat---directly into a voltage through the thermoelectric effect.\cite{Innes1985,Weeber2011}
We employ two local gates to fully tune the thermoelectric and plasmonic behaviour of the graphene.
High-resolution real-space photocurrent maps are used to investigate the plasmon propagation and interference, decay, thermal diffusion, and thermoelectric generation.
}


Graphene plasmonics is an emerging platform for terahertz to infrared nano-optics, attractive due to the long intrinsic lifetime of $> 0.5$~ps and the strong tunable broadband electrodynamic response of its Dirac electrons.\cite{Jablan2009,Principi2014}
Typically, graphene plasmons are sensed by out-coupling to light, which is inefficient due to one of the key features of graphene plasmons: their extremely short wavelength ($\sim \tfrac{1}{100}$ that of free space light).
While plasmon resonances have been exploited to enhance absorption and thereby enhance far-field photodetection,\cite{Freitag2013,Cai2015} the concept of an on-chip plasmon receiver has not yet been realized.

The presented experimental device is built around the state-of-the-art plasmonic medium of graphene encapsulated in hexagonal boron nitride (hBN), which we have recently demonstrated to support high quality propagating plasmons in the mid infrared.\cite{Woessner2014}
As a key innovation in this work, the induction of free carriers in the graphene is achieved through the use of separate local gates directly underneath the hBN, rather than the conventional global back gating through an additional SiO$_2$ layer.
The use of local gating allows to spatially modulate the charge carrier density and polarity across the device, as well as providing lower voltage operation and reduced charge trapping effects.\cite{Woessner2015}
As we will show below, the junction induced by the two gates can be used as a thermoelectric detector for the plasmons.

\begin{figure}[t]
\includegraphics{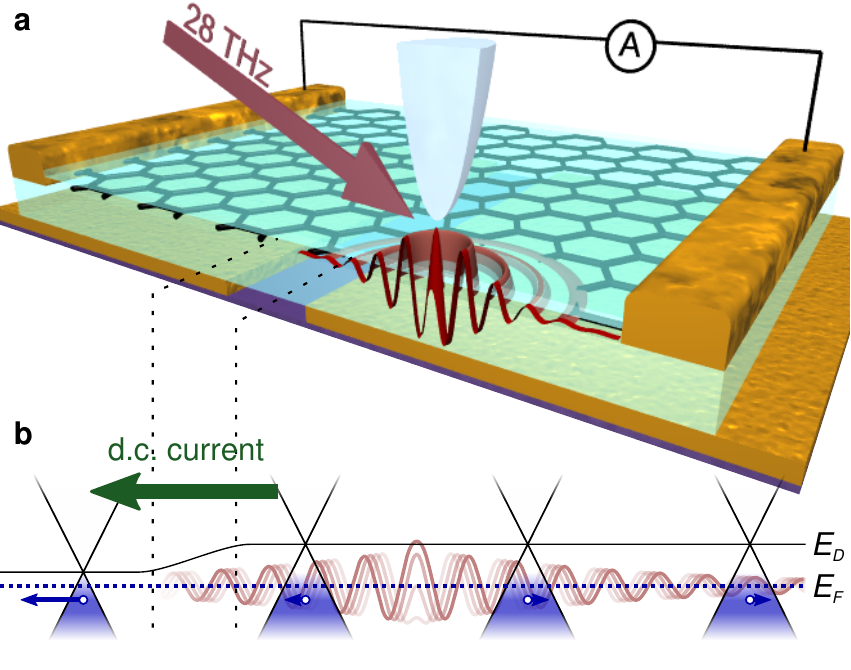}

\includegraphics{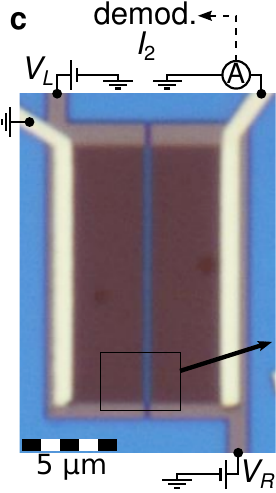} \hfill 
\includegraphics{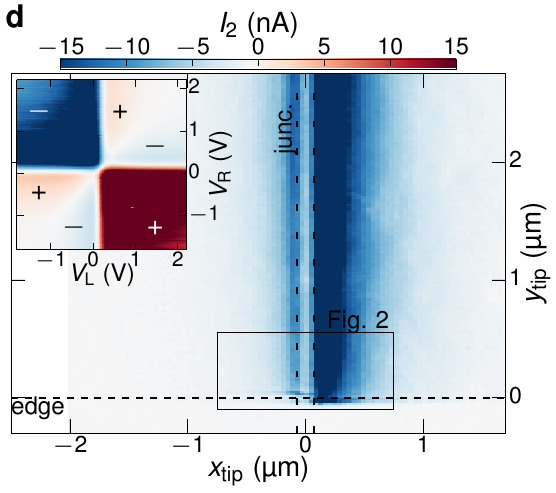}

\caption{
\label{fig1}
\textbf{\em Concept and device.}
\textbf{a},
Schematic cross-section of device and measurement technique.
Continuous-wave laser light scatters at a movable metallized AFM tip, launching plasmons in the hBN--graphene--hBN heterostructure.
\textbf{b},
Schematic of thermoelectric detection mechanism in a microscopic picture.
The plasmon decay energy drives an outward majority carrier diffusion, in this case hole carriers.
A gate-induced homojunction (seen as variation in the graphene Dirac point energy level $E_{\rm D}$ relative to the Fermi level $E_{\rm F}$) imbalances this diffusion, resulting in a nonzero net dc current.
\textbf{c},
Optical micrograph of presented device and circuit diagram.
Two metal electrodes (light yellow) contact an encapsulated graphene sheet (dark rectangle) which lies above a split metal gate layer (light brown).
Split gate voltages $V_{\rm L}$ and $V_{\rm R}$ create the homojunction, while tip-induced currents are captured at the electrodes and demodulated to obtain the near field photocurrent $I_2$.
\textbf{d},
Near-field photocurrent map of the entire device, showing the photosensitive junction created by applying different gate voltages.
{\em Inset}: The sign of the photocurrent (measured with the tip over the junction) shows a six-fold pattern characteristic of thermoelectric effects.
}
\end{figure}

Figures \ref{fig1}a,b show a schematic of the operating principle of the detector.
In lieu of an on-chip plasmon source, we generate plasmons using the conventional scattering scanning near field microscopy (s-SNOM) technique.\cite{Fei2012,Chen2012}
The s-SNOM apparatus consists of a scanning metal probe under illumination from a continuous wave laser at mid infrared frequency.
A laser frequency of 28~THz (\SI{10.6}{\micro\metre} free space wavelength) was chosen to avoid complications from the hBN phonons.\cite{Woessner2014}
In conventional plasmonic s-SNOM experiments, the signal of interest is the out-scattered light, containing information about local dielectric properties and plasmonic modes.
Here, we instead measure a quantity $I_2$, known as {\em near field photocurrent}, from the current exiting the device electrodes (Fig.~\ref{fig1}c).\cite{Woessner2015}
This is the component of total measured current that oscillates at the second harmonic ($\sim$500~kHz) of the probe tapping frequency ($\sim$250~kHz).
As the graphene shows a linear photocurrent response, $I_2$ can be understood as the photocurrent arising only from the $\sim$60~nm-sharp near fields of the tip, isolated from the background photocurrent directly induced by the incident light.
For simplicity, in the remainder of this paper we refer to $I_2$ simply as ``the photocurrent'' and treat it as if it arises from an effective nanoscale light source.

The studied device and circuit schematic is shown in Fig.~\ref{fig1}c.
By applying different voltages $V_{\rm L(R)}$ to the left~(right) gates, we induce a localized photosensitive region, e.g., a $p$--$n$ junction as studied in Fig.~\ref{fig1}d.
The six-fold photocurrent pattern observed when both gates are scanned (figure inset) is considered as evidence of a thermoelectric generation mechanism, where the pattern arises due to the nonmonotonic dependence of Seebeck coefficient on gate voltage.\cite{Xu2010,Lemme2011,Song2011,Gabor2011}
For a simple junction, the thermoelectric current is $I = (S_{\rm R} - S_{\rm L}) \overline{\Delta T}^{\rm junc} / R$, where $S_{\rm L(R)}$ is the left~(right) Seebeck coefficient, $\overline{\Delta T}^{\rm junc}$ is the junction-average rise in electronic temperature relative to ambient, and $R$ is the circuit resistance.
The gate dependence allows to identify the charge neutrality point of the graphene in this device (occuring at a gate voltage offset of +0.09~V).
Hereafter we use this offset and the calculated gate capacitance to convert the gate voltages $V_{\rm L,R}$ into carrier densities $n_{\rm L,R}$.

\begin{figure}[t]
\includegraphics{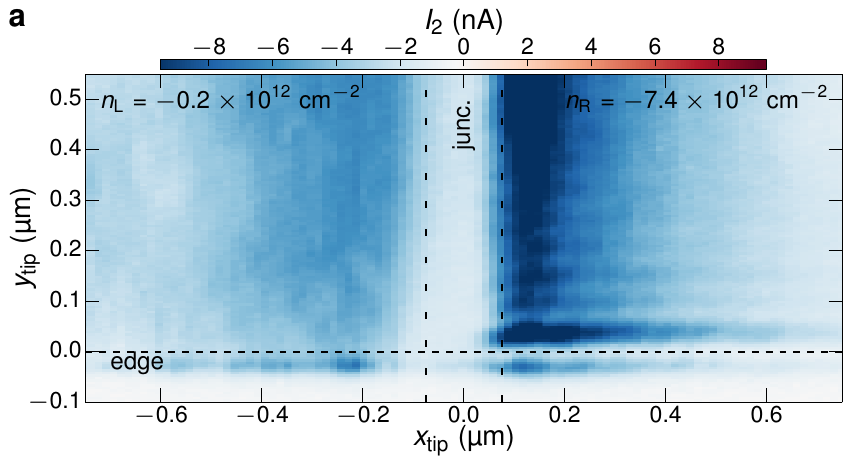}
\includegraphics{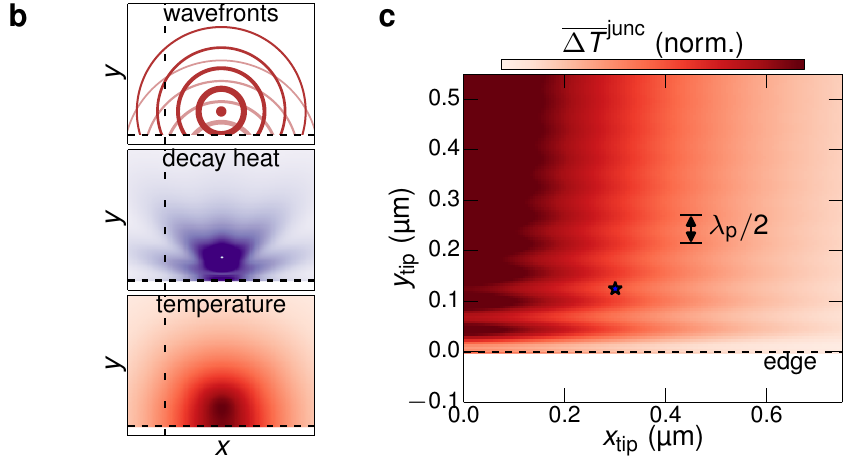}
\caption{
\label{fig2}
\textbf{\em Plasmon photocurrent spatial maps.}
\textbf{a},
A high-resolution photocurrent map near the edge of the graphene, containing interference fringes.
\textbf{b},
Modelled fields for a given $x_{\rm tip}$, $y_{\rm tip}$ position: wave propagation and interference (upper panel: strong red curves are launched wavefronts, faint red curves are reflected wavefronts), decay to heat (middle panel), and thermal spreading (bottom panel).
Parameters used were $k_{\rm p} = (56 + 1.8i)~\si{\micro\meter^{-1}}$, $r = 0.4 e^{0.65\pi i}$, $l_{T} = \SI{0.25}{\micro\meter}$.
\textbf{c},
The modelled average temperature rise along the junction (along the vertical dashed line in (b)), $\overline{\Delta T}^{\rm junc}$, and its dependence on $x_{\rm tip}$, $y_{\rm tip}$.
The $\star$ symbol marks the case shown in panel {\bf b} of this figure.
}
\end{figure}

Strong evidence of plasmons mediating the photocurrent is visible in photocurrent maps obtained at high carrier density (Fig.~\ref{fig2}a), where interference fringes can be observed in $I_2$ near the graphene edge.
These fringes can be unambiguously attributed to plasmon reflections, as they match the half-wavelength periodicity seen in the s-SNOM optical signal that is conventionally used to characterize graphene plasmons.\cite{Woessner2014}
The extracted plasmon wavelength of $\lambda_{\rm p} = 112$~nm in this scan is close to the expected value of 114~nm, and consistent with a previous study of a similar hBN--graphene--hBN device.\cite{Woessner2014}

To explain the spatial $I_2$ pattern and the detection mechanism, we consider the following sequence, sketched in Fig.~\ref{fig2}b:
Plasmons radiate away from the tip and reflect at the edge;
the self-interfered plasmon wave decays into electronic heat;
subsequent electronic diffusion spreads the heat to the junction, determining $\overline{\Delta T}^{\rm junc}$.
To justify this interpretation, we employ a simplified two-dimensional model that takes into account each of these effects.
The model, detailed in Methods, yields the value of $\overline{\Delta T}^{\rm junc}$ (up to a normalization) for a plasmon source located at any position $x_{\rm tip}$, $y_{\rm tip}$.
The three critical model parameters are plasmon wavevector $k_{\rm p}$, reflection coefficient $r$, and electron cooling length $l_{T}$.
By varying these, we obtain a map (Fig.~\ref{fig2}c) that fits to the data, capturing the essential physics behind the observed spatial pattern.
Note that this model neglects direct three-dimensional near field coupling effects, giving some disagreement within $\sim 100$~nm of the edge and junction.

\begin{figure}[t]
\includegraphics{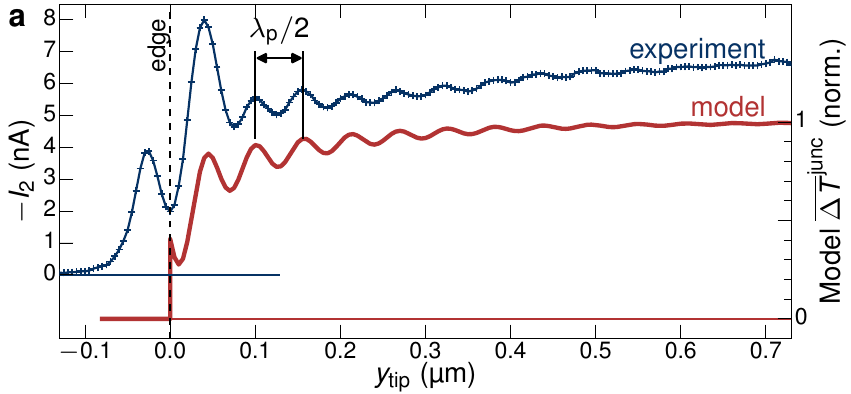}
\includegraphics{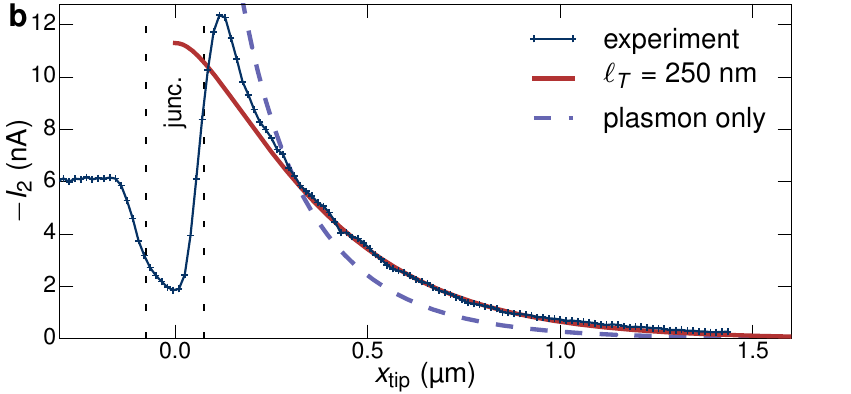}
\caption{
\label{fig3}
\textbf{\em Linecuts along $x_{\rm tip}$ and $y_{\rm tip}$.}
\textbf{a},
Linecut of photocurrent map perpendicular to the edge, obtained by averaging the data in Fig.~\ref{fig2}a over the interval $x_{\rm tip} = 0.2 \cdots \SI{0.4}{\micro\metre}$.
The lower curve (right axis) shows the corresponding model linecut (from Fig.~\ref{fig2}c).
\textbf{b},
Linecut of photocurrent across the junction, far from the edge, obtained by averaging Fig.~\ref{fig2}a over the interval $y_{\rm tip} = 0.6 \cdots \SI{0.7}{\micro\metre}$.
The solid red curve shows the corresponding model linecut (from Fig.~\ref{fig2}c), and the dashed curve shows the model result considering only plasmon propagation (without thermal diffusion); model curves have been vertically scaled for comparison with the data.
}
\end{figure}

Two complex parameters, $k_{\rm p}$ and $r$, are key for matching the $y_{\rm tip}$ dependence, examined in detail in Fig.~\ref{fig3}a.
Whereas $\Re[k_{\rm p}] = 2\pi/ \lambda_{\rm p}$ determines the fringe spacing, $\Im[k_{\rm p}]$ encodes the plasmon decay length and determines the number of visible fringes.
In particular, the fringes decay according to an envelope function $\exp(-y_{\rm tip} \Im[k_{\rm p}])/\sqrt{y_{\rm tip}}$, identical to that of the optical signal.\cite{Woessner2014}
The reflection coefficient $r$ is relevant for setting the overall magnitude and phase of the fringes, from $|r|$ and $\operatorname{arg}(r)$ respectively.
The subunity value of $|r| = 0.4$ also leads to a drop in power as the tip is brought near the edge, since in this model the unreflected plasmon power is lost.
A similar drop is seen in the data, suggesting that the unreflected plasmon power is not converted to electronic heat in the same way as for plasmon decay elsewhere.

The electron cooling length, $l_{T}$, is important for matching the photocurrent decay away from the junction, shown in Fig.~\ref{fig3}b.
This $l_{T}$ is the typical distance of electronic thermal diffusion before the heat is conducted out of the electronic system, and hence correponds to the effective length over which the junction is sensitive to heat inputs (in this case, plasmon decay heat).
At this point it is worthwhile to compare to other hypothetical non-thermal detection mechanisms, where the junction would sense directly the incident plasmon power.
In that case, the signal would be proportional to the average plasmon intensity precisely at the junction, and hence proportional to the un-diffused decay heat along the junction.
As we show in Fig.~\ref{fig3}b, such mechanisms would produce a too-short decay length, determined only by the plasmon energy propagation length, $\Im[2 k_{\rm p}]^{-1}$.

\begin{figure}[t]
\includegraphics{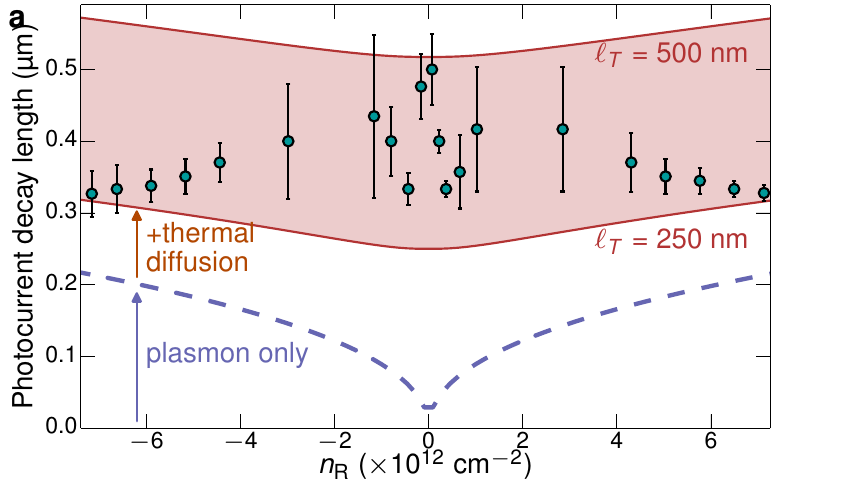}
\includegraphics{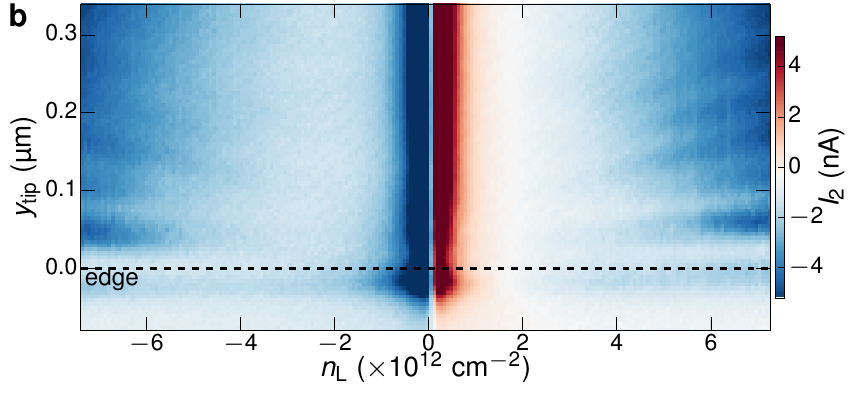}
\includegraphics{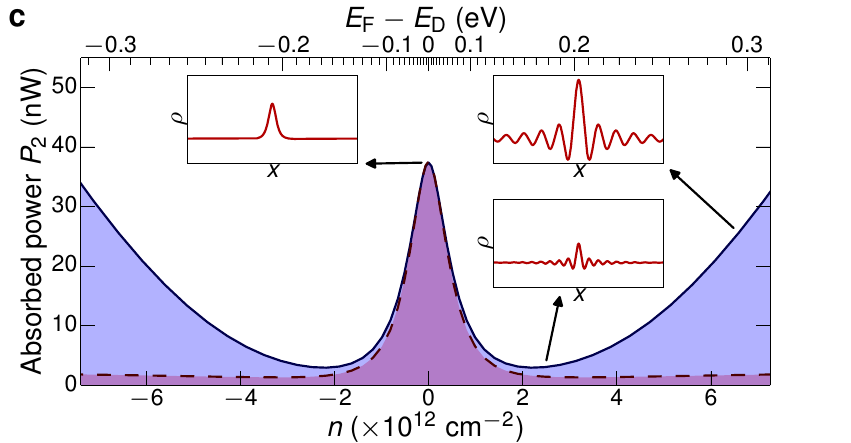}
\caption{
\label{fig4}
\textbf{\em Gate dependence.}
\textbf{a},
Carrier density dependence of photocurrent decay length away from junction.
The decay length was obtained by an exponential fit to $I_2(x_{\rm tip})$ for  $x_{\rm tip} > \SI{0.3}{\micro\meter}$, far away from the edge.
This was done for several values of $n_{\rm L} \sim -2 \cdots \SI{+2e12}{cm^{-2}} $, resulting in the error ranges shown.
The solid curves show the corresponding decay in our model assuming the indicated thermal lengths, and the dashed curve shows the result neglecting thermal diffusion.
\textbf{b},
Dependence of photocurrent on $n_{\rm L}$, at various tip positions away from the graphene edge.
This scan was taken 300~nm left of the junction with $n_{\rm R} = \SI{0.26e12}{cm^{-2}}$.
\textbf{c},
Power absorbed in graphene, calculated from a rounded-tip electrodynamic model.
The dashed curve shows the absorbed power as it would be with only the real part of the graphene conductivity retained (i.e., without plasmons).
{\em Insets}: The induced charge density oscillation in the graphene calculated for various carrier densities.
}
\end{figure}

The requirement of some diffusion to match the data confirms our picture that the detection mechanism does not rely on direct rectification of the plasmon at the junction, but instead is based on sensing the temperature rise from plasmon decay.
Further evidence along this line is shown in Fig.~\ref{fig4}a, where we have analyzed the photocurrent decay by a fitted exponential decay length, at several different carrier densities.
This dependence disagrees both quantitatively and qualitatively with a direct detection mechanism.
Instead, a density-dependent value of $l_{T}$ from $\sim 500$~nm (low $|n|$) to $\sim 250$~nm (high $|n|$) is needed.

Next, we show tunability of the nature and strength of the plasmon launching, with varying carrier density (Figs.~\ref{fig4}b,c).
Figure \ref{fig4}b shows the dependence of the photocurrent on the gate voltage under the tip.
The data show several features simultaneously evolving with carrier density.
There are two sign changes in photocurrent, due to the sign change of Seebeck coefficient differences.\cite{Xu2010,Lemme2011,Song2011,Gabor2011}
The fringe spacing appears to follow $\tfrac{1}{2}\lambda_{\rm p} \propto |n|^{1/2}$ as expected for graphene plasmons.\cite{Woessner2014}
Most strikingly, the photocurrent shows two regimes of strong magnitude, at high $|n|$ and low $|n|$, separated by a region of weak photocurrent from $|n| \sim 1$--$4\times 10^{12}~\mathrm{cm^{-2}}$.
We attribute these to the two ways that graphene can absorb power from the tip: direct heating or plasmon launching, which are both captured in our quantitative electrodynamic calculations of the absorbed power (Fig.~\ref{fig4}c, details in Methods).
The launched plasmon power grows strongly with carrier density primarily due to variation in plasmon wavelength: plasmons with small values of $\lambda_{\rm p}$ couple poorly to the tip due to their strong confinement in the top hBN layer and the limited range of spatial frequencies probed by the round tip.
Direct heating on the other hand is strongest for low $|n|$ due to unblocked interband transitions, possible when the Dirac point $E_{\rm D}$ is within about $\hbar\omega/2$ ($= 58$~meV) of the Fermi level $E_{\rm F}$.

The observed photocurrent signals are well above the noise level, and according to the calculation of Fig.~\ref{fig4}c these signals arise from a plasmon power of order 10~nW---note that there is some uncertainty in this number due to the difficulty of accurately modelling the tip.
This order-of-magnitude power estimate can be compared to that predicted for local plasmon sources, to see whether such compact sources could serve in place of our plasmon-launching tip.
One proposed plasmon source is the graphene thermal plasmon radiator,\cite{Liu2014} which is a hot graphene strip (at $\sim 500$~K) adjacent to a room temperature graphene channel.
Such a source would emit plasmon power on the order of tens of nanowatts,\cite{Liu2014} which should be detectable using our junction device.

In conclusion, we have shown that a graphene homojunction serves as an electrical detector for the mid-infrared plasmons that are carried by the graphene itself.
The available evidence strongly indicates that thermoelectric action is detecting the energy of the plasmon after it has decayed and that thermal diffusion plays an important role in spreading the decay energy.
The presented concept opens the door to graphene plasmonic devices where inefficient plasmon out-coupling to light is unnecessary.
We anticipate in the future that this detector may be paired with a local plasmon source such as those based on thermal\cite{Liu2014} or tunneling emission,\cite{Svintsov2015} resulting in an end-to-end mid infrared optical system at sizes far below the light diffraction limit.

\section*{Methods}

{\small
Device fabrication started with an 10~nm, surface low roughness AuPd alloy gate film patterned by electron beam lithography, on top of an oxidized Si substrate.
The gap separating the gates from each other was 150~nm, as indicated in the figures.
An hBN--graphene--hBN stack was then prepared by the van der Waals assembly technique,\cite{Wang2013} and placed on top of the AuPd gate layer.
The bottom hBN film (between graphene and metal) thickness of 27~nm was chosen to isolate the plasmonic mode from interacting with the gate metal, while still allowing for strong gate effect.
The top hBN film was made thin (9~nm) to allow for plasmon launching by the s-SNOM method.
The device geometry as well as the edge contacts were defined by dry etching and electron beam evaporation in the method of Ref.~\onlinecite{Wang2013}.
The dry etching depth was only 11~nm, leaving most of the bottom hBN thickness remaining in order to avoid leakage.
Gate voltages were converted to carrier sheet density via $n_{\rm L,R} = (\SI{0.73e16}{m^{-2} V^{-1}} )(V_{\rm L,R} - 0.09~\mathrm{V})$, where the offset was determined by examining gate dependences and the coefficient was calculated as the static capacitance of the 27~nm hBN layer with dielectric constant 3.56.\cite{Woessner2014}

The s-SNOM used was a NeaSNOM from Neaspec GmbH, equipped with a CO$_2$ laser.
The probes were commercially-available metallized atomic force microscopy probes with an apex radius of approximately 25~nm.
The tip height was modulated at a frequency of approximately $250~\mathrm{kHz}$ with amplitude of 60~nm.
The location of the etched graphene edge ($x_{\rm tip}=0$) was determined from the simultaneously-measured topography.\cite{Woessner2014}

In Fig.~\ref{fig2}, we solve the Helmholtz wave equation
\begin{equation}
 k_{\rm p}^{2} \rho(x,y) + \nabla^2 \rho(x,y) = f(x,y),
\label{eq:helmholtz}
\end{equation}
for a localized sourcing distribution $f(x,y)$ (concentrated at $x_{\rm tip}$, $y_{\rm tip}$), where $k_{\rm p}$ is the complex plasmon wavevector.
Here $\rho$ represents the spatial dependence of the oscillating charge density $\Re[\rho(x,y)e^{-i\omega t}]$.
The reflective boundary at $y=0$ is asserted by the method of images: solving \eqref{eq:helmholtz} for free space, adding a virtual copy at $-y_{\rm tip}$ multiplied by a complex reflection coefficient $r$, and discarding the virtual solution below $y=0$.
Dissipation in the graphene converts the plasmon to a decay heat distribution proportional to $|\vec\nabla \rho|^2$.
This heating distribution is diffused,
\begin{equation}
 l_{T}^{-2}(T(x,y) - T_0) - \nabla^2 T(x,y) \propto |\vec\nabla \rho(x,y)|^2
\end{equation}
to yield $T(x,y)$, the local temperature distribution, with edge boundary condition $\partial_y T |_{y = 0} = 0$.
The parameter $l_{T}$ is the characteristic length of lateral heat spreading before sinking to the substrate at temperature $T_0$.
Finally the average temperature rise on the junction, which drives the thermoelectric effect, is represented by the quantity $\overline{\Delta T}^{\rm junc}$:
\begin{equation}
 \overline{\Delta T}^{\rm junc} = \frac{1}{W}\int_0^W \d y T(0,y) - T_0
\end{equation}
for device width $W$, and it is this quantity plotted in Fig.~\ref{fig2}c.
The value of $W$, strength of $f(x,y)$, and other proportionality factors drop out due to normalization.
The case of direct plasmon detection is found in the limit $l_{T} \rightarrow 0$, in which case the signal is determined by the $y$-integral of $|\vec\nabla \rho|^2$.

The solid curves in Figure \ref{fig4}a result from performing an exponential fit to the modelled $\overline{\Delta T}^{\rm junc}$.
For each $|n|$ we estimated $k_{\rm p}$ using the fitted $k_{\rm p}$ from high $|n|$ (Figs.~\ref{fig2},\ref{fig3}) and the trend $1/k_{\rm p} \propto \sqrt{|n|}$ found in our previous study.\cite{Woessner2014}

Our electrodynamic calculation (Fig.~\ref{fig4}c) consists of a tip charge distribution, calculated via a regularized boundary-element electrostatic model,\cite{McLeod2014} fed into a multilayer transfer matrix calculation for the hBN--graphene--hBN-metal stack.
An incident field of 0.3~MV/m was estimated from the experimental 10~mW incident laser power, which is focussed to a diffraction-limited spot (NA 0.5, \SI{10.6}{\micro\meter} wavelength).
The tip surface was taken as a circular hyperboloid of $50^{\circ}$ opening angle and a 25~nm curvature radius at the apex, with a \SI{5}{\micro\meter} length yielding a 45$\times$ tip electric field enhancement factor over the incident field.
The 3D charge distribution was remapped to a 2D charge distribution located a distance $z_{\rm tip}$ from the top hBN surface and this distribution was oscillated at 28~THz, with accompanying in-plane currents.
The absorbed power in the graphene, $\frac{1}{2} \Re[\vec J^* \cdot \vec E]$, was calculated for 36 different $z_{\rm tip}$ values from 0 to 60~nm, and this height dependence was then used to simulate the second harmonic demodulation process, arriving at a second-harmonic power $P_2$ that best corresponds to the studied current $I_2$.
The hBN relative permittivity at this frequency was taken as $8.27+0.16i$ in-plane and $1.88+0.04i$ out-of-plane.\cite{Woessner2014}
The graphene conductivity used was the local finite-temperature RPA conductivity formula,\cite{Falkovsky2007} taking care to map from $(E_F-E_D)$ to $n$ using the appropriate Fermi-Dirac integral.

}


\begin{acknowledgments}
{ \small
We thank Marco Polini, Alexey Nikitin, and Klaas-Jan Tielrooij for fruitful discussions.
This work used open source software (www.matplotlib.org, www.python.org, www.povray.org).
F.H.L.K. and R.H. acknowledge support by the EC under Graphene Flagship (contract no.\ CNECT-ICT-604391).
F.H.L.K. acknowledges support by Fundacio Cellex Barcelona, the ERC Career integration grant (294056, GRANOP), the ERC starting grant (307806, CarbonLight), the Government of Catalonia trough the SGR grant (2014-SGR-1535), the Mineco grants Ramón y Cajal (RYC-2012-12281) and Plan Nacional (FIS2013-47161-P).
P.A.-G. and R.H. acknowledge support from the European Union through ERC starting grant (TERATOMO grant no. 258461) and the Spanish Ministry of Economy and Competitiveness (national project MAT2012-36580).
Y.G., C.T., and J.H. acknowledge support from the US Office of Naval Research N00014-13-1-0662.
} \end{acknowledgments}

\bibliography{bib-jabref}

\end{document}